\documentclass[12pt]{article}
\usepackage{amsmath}
\usepackage{eufrak}
\usepackage{braket}
\usepackage{authblk}
\usepackage{amssymb} 
\usepackage{graphicx}
\usepackage{setspace}

\newcommand{\beq}{\begin{equation}}
\newcommand{\eeq}{\end{equation}}
\providecommand{\keywords}[1]
{\small	
  \textbf{\textit{Keywords---}} #1}

\begin{document}
\onehalfspacing
\begin{center}
{\Large Sum of Three Squares and Particle in a Cubic Box -  Degeneracies}

\vspace{0.5cm}

{\large S. Pratik Khastgir}\footnote[2]{pratik@phy.iitkgp.ac.in}

\vspace{0.5cm}

{  Department of Physics, Indian Institute of Technology Kharagpur, Kharagpur 721302, India }

\end{center}
\hspace{1cm}\keywords{Three square theorem, Particle in a box, Eigenvalues,  Degeneracy}

\begin{abstract}
We address the problem of determining the degeneracy of a particular energy eigenvalue corresponding to a quantum particle trapped in a cubic box. In other words, we aim to find the number of ways a specific positive integer can be written as the sum of three natural number squares. This is a classic problem in number theory. We provide a closed-form, exact solution to the problem. We do not claim to provide proofs, as we have none; instead, we build up the formulae starting from simple cases where the answer could be deduced by studying the list of degeneracies. We present ample examples that could be verified against the degeneracy calculated numerically via the brute-force method. We arrived at the solution by meticulously studying the list of degeneracies for the first five million numbers, generated by Mathematica\textsuperscript{\textregistered}.

\end{abstract}

\section{Introduction} The story begins  with the ways of expressing a positive integer $n$ as a sum of $\kappa$ integer squares \cite{Dickson_1920,Grosswald_1985}. That is, writing
\beq
n=\sum_{i=1}^{\kappa}a_i^2; \qquad a_i\in \mathbb{Z}
\label{progen}
\eeq
with distinct sets of $a_i$s. The total number of ways of representing $n$  (that is, number of sets of $a_i$s, including positive, negative and zero values for $a_i$s) is denoted by the symbol $r_\kappa (n)$.  The problem becomes more involved when one restricts $a_i$s to, for instance,  only non-negative integers, or only natural numbers  
\cite{Grosswald_1985,Grosswald-etal_1959, Baltes-etal_1974}. If $a_i\in \mathbb{N}$  then the total number of ways is denoted by $c_{\kappa}(n)$. Note, in $c_{\kappa}(n)$ we count the permutations of $a_i$ as different solutions. For example in $\kappa=2$ case, $a_1=a$, $a_2=b$  and $a_1=b$, $a_2=a$ are counted as two different solutions. If we know $c_{\kappa}(n)$ we can calculate  $r_\kappa (n)$. For $\kappa=1$ the problem is simple, $c_1(n)=1$ when $n$ is a square number, that is, $n=a_1^2, a_1\in \mathbb{N}$ and $c_1(n)=0$ otherwise. Since, $n=(\pm a_1)^2$ in this case,  $r_1(n)=2c_1(n)$. In $\kappa=2$ case, if $n=(\pm a_1)^2+(\pm a_2)^2$ where $a_1, a_2\in  \mathbb{N}$, then $r_2(n)=4c_2(n)+2r_1(n)=4c_2(n)+4c_1(n)$.  In the previous expression second term appears for the case when either of $a_1$ or $a_2$ is zero. It was realised soon that the $\kappa=3$ problem is considerably harder than the $\kappa=2$ and $\kappa=4$ problems \cite{Bateman_1951, Grosswald_1985}.  With some algebra one can show that \cite{Mahlburg_2004},  
\beq
r_3(n)=8c_3(n)+12c_2(n)+6c_1(n).
\label{count}
\eeq
In the following we concentrate on the $\kappa=3$ problem.
The three square theorem, credited to Gauss and Legendre  \cite{Grosswald_1985}, states that $r_3(4^j(8k+7))=0;~ j,k=0,1,2,\cdots,$  that is if $n$ is of the form $4^j(8k+7)$ then it cannot be represented as a sum of three squares. For  $n>4$ and squarefree, Gauss could write  $r_3(n)$ in terms of the class number, $h(-n)$, of imaginary quadratic  field  $\mathbb{Q}(\sqrt{-n})$  \cite{Dickson_1920,Grosswald_1985,Lemmermeyer_2021}. For non-squarefree integers, $n$, the problem becomes complex,  Bateman provided the answer for $r_3(n)$ in terms of the Dirichlet L-function, \cite{Bateman_1951}.  Useful results are available for $r_3(n)$ for  specific choices of $n$, \cite{Hirschhorn-etal_1999}. A new proof of Gauss formula of $r_3(n)$ is presented in reference \cite{Mortenson_2017}.   The situation, $a_i\in \mathbb{N}$, naturally arises in elementary Quantum Mechanics while tackling the problem of a quantum particle of mass $\mu$  trapped in a cubic box
of length $L$. By solving the Schr\"odinger equation with proper boundary conditions we get the energy spectrum \cite{Pauling_1935}:,
\beq
{\cal E}=\frac{\hbar^2\pi^2}{2\mu L^2}(a^2+b^2+c^2); \qquad a,b,c \in  \mathbb{N}
\label{en0}
\eeq
Without loss of generality the combination of parameters $\frac{\hbar^2\pi^2}{2\mu L^2}$ could be set to 1. The problem, in its new avatar, is now identical to determining the degeneracy, ${\cal {\cal D}}({\cal E})=c_3({\cal E})$, of the energy state ${\cal E}$. It could be seen easily that, 

${\cal {\cal D}}({\cal E}=3)=1;~ (a=1,b=1,c=1),$ 

${\cal {\cal D}}({\cal E}=6)=3; ~(a=1,b=1,c=2; a=1,b=2,c=1;a=2,b=1,c=1), $

${\cal {\cal D}}({\cal E}=14)=6;~ (a=1,b=2,c=3;~ {\rm and~ permutations}). $

\noindent The degeneracies appearing in ${\cal E}=6$ or  ${\cal E}=14$ are
{\it expected degeneracies} since they are inherited from the symmetry of the system. 
For a cubic box all the three directions are equivalent, there is nothing which could distinguish one from the other. In this case one could exchange $x$ with $y$ or $y$ with $z$ or $z$ with $x$ in a particular eigen function and if that results in a new eigen function then the corresponding  energy would remain the same. This exchange is achieved by permuting the quantum numbers $a, b$ and $c$ in the expression of energy  ${\cal E}$. So in a set of ($a$, $b$, $c$) when two numbers are the same it gives a degeneracy of 3 whereas all three different results in a degeneracy of 6. This count is easy. The difficulty arises due to {\it accidental degeneracy}, where the symmetry is not apparent but there is a degeneracy due to the particular dependence on quantum numbers in the expression of ${\cal E}$. For example ${\cal E}=27$ could be written as $3^2+3^2+3^2$ or as $1^2+1^2+5^2$. The former contributes 1 set whereas the latter contributes 3, so the total degeneracy of the state  ${\cal E}=27$ is $1+3=4$. Let us take another example, ${\cal E}=41=1^2+2^2+6^2=3^2+4^2+4^2$, the first combination contributes  6 and the second 3 to make the total degeneracy 9 of this state. So, the actual count of degeneracy boils down to identifying the distinct sets of natural numbers $(a,b,c)$ with any two of them same and all three different. The case $a=b=c$ will just contribute an extra 1 to the degeneracy count. This case is easy to recognise as it will be fulfilled when ${\cal E}/3$ is a square number. To the best of our knowledge counting ${\cal D}{(\cal E)}$ (that is  $c_3({\cal E}$)) for a given positive integer ${\cal E}$ is not yet solved completely \cite{Grosswald_1985}.  In the present note we try to fill this gap.
 All the results available for $r_3(n)$ are involved and are not easy to interpret. We present formulae to calculate $c_3(n)$ in terms of factors of $n$  and their powers.
Our formulae can be used to calculate $c_3(n)$ for large $n$ efficiently and in turn could be exploited to calculate $r_3(n)$ as well, using the formula (\ref{count}) given in the beginning.

In the next section we briefly present the results of $\kappa=2$ case with a variation on the theme. Next we move to the $\kappa=3$ case where we provide the full solution of the problem considering the different cases explicitly. Our results show that in half the cases (where prime factorisation of ${\cal E}$ results only in even powers of odd rational primes), the degeneracy could be obtained without involving the class number, $h(-d)$, ($d\in \mathbb{N}$ and squarefree) of the  imaginary quadratic field, $\mathbb{Q}(\sqrt{-d})$, whereas the other half (where odd powers of rational primes are involved in the factorisation of ${\cal E}$) would require the involvement of class numbers. We stress again that we construct formulae based on observations, that is, by thoroughly  studying the degeneracy list created numerically through brute force.  We substantiate our formulae through numerous examples. In the following ${\cal E}$ is synonymous to $n$.

\section{Particle in a two-dimensional box} 

We present this problem as a warm-up exercise. This section is a summary of the results presented in reference \cite{Shaw_1974}. Our contribution is the unified formula. A quantum particle of mass $\mu$ trapped in a two dimensional square box of length $L$ is exactly solvable. With the separation of variables, the solution of Schr\"odinger equation with proper boundary conditions results in the energy spectrum \cite{Park_1974},
\beq
{\cal E}=\frac{\hbar^2\pi^2}{2\mu L^2}(a^2+b^2); \qquad a,b \in  \mathbb{N}
\label{en1}
\eeq
 Without loss of generality, setting  $\frac{\hbar^2\pi^2}{2\mu L^2}=1$ reduces the problem to the one described in (\ref{progen}) with $\kappa=2$ and $a_i$s restricted to natural numbers. 

Another exactly solvable case is when the particle is trapped in a two dimensional equilateral triangular box of side length $L$ \cite{Mathews_1970}. In this case the spectrum is given by
\beq
{\cal E}=\frac{8\hbar^2\pi^2}{9\mu L^2}(a^2+ab+b^2); \qquad a,b \in  \mathbb{N}
\label{en2}
\eeq
In this case too without loss of generality one could set $\frac{8\hbar^2\pi^2}{9\mu L^2}=1$. It turns out that  the question of degeneracy of a particular energy level ${\cal E}$ for the two cases above, (\ref{en1}) and (\ref{en2}) can be answered with a single expression. From the bilinear structure it is easily seen that the degeneracy ${\cal D}^{(2)}({\cal E'}={4\cal E})={\cal D}^{(2)}({\cal E}).$ This is achieved by just substituting $a'=2a$ and $b'=2b$ to construct ${\cal E'}=a'^2+b'^2$ and ${\cal E'}=a'^2+a'b'+b'^2$ when $a$ and $b$ satisfy
${\cal E}=a^2+b^2$ and ${\cal E}=a^2+ab+b^2$ respectively. On the other hand, when  ${\cal E'}=a'^2+b'^2$ or ${\cal E'}=a'^2+a'b'+b'^2$ and ${\cal E'}={4\cal E}$
both $a'$ and $b'$ must be even and we can write  $a'=2a$ and $b'=2b$, so there is a one to one correspondence between the sets  $(a,b)$ and $(a',b')$.
Let us start with the expression of viable energy in these two cases:

\subsection{Square box, $\kappa=2$}
 Taking all the possibilities:
\begin{enumerate}
\item{{\it $ a, b, $ both even }: $a=2l, ~b=2m$,  then ${\cal E}=4(l^2+m^2)$ which is  0 modulo 4. }

\item{{\it $ a, b, $ one even and one odd} : $a=2l, ~b=2m+1$,  then ${\cal E}=4[l^2+m(m+1)]+1$  which is  1 modulo 4. }

\item{{\it $ a, b,$ both odd} : $a=2l+1, ~b=2m+1$,  then ${\cal E}=4[l(l+1)+m(m+1)]+2=2(4k+1)$ since products of two consecutive integers like $l(l+1)$ are always even and we may write $[l(l+1)+m(m+1)]=2k$. }

\end{enumerate}

From the above we conclude that the viable energies in this case are of the type  ${\cal E}=2^\beta(4k+1)$, $k,\beta=0,1,2\cdots$. Note that not all the combinations of $k$ and $\beta$ are allowed.  Stripping all the powers of $2$ off we are left with  $4k+1$.  The above also proves that the sum of two squares cannot be of the type $4k'+3$ or $4k-1(k=k'+1)$ type. The viable energies for a particle trapped in square box could be written into prime factorisation \cite{Grosswald_1985, Shaw_1974}:
 \beq
{\cal E}=2^{\beta}\prod_{i=1}^{I}(4l_i-1)^{2\lambda_i}\prod_{j=1}^{J}(4m_j+1)^{\nu_j},
\label{en3}
\eeq
where the exponents, $\beta,\lambda_i$ and $\nu_j$ are non-negative integers. $l_i$ and $m_j$ are positive integers such that $4l_i-1$ and $4m_j+1$ are rational odd primes.
The  $4m_j+1$ type primes  can have any power even or odd since the product of $4m_j+1$ type primes will always result in a number of the $4k+1$ type. Note that the $4l_i-1$ type primes have only the even powers. It is easy to see that sum of powers of $4l_i-1$ type primes cannot be odd, since the product of odd number of $(4l_i-1)$ type primes will result in a number of the type  $4k-1$ and which cannot be written as $a^2+b^2$ as shown above. 

\subsection{Equilateral triangular box}
For a particle  trapped in an equilateral triangular box, like the previous case when both $a$ and $b$ are even we have ${\cal E}=4k$. When both $a$ and $b$ are odd or when one of them is even and the other is odd, a little algebra shows that ${\cal E}=4k+1$ or  ${\cal E}=4k+3$.  So the energies, ${\cal E}=4k+2$ are not viable in this case. This in turn says that the odd powers of 2 are absent in the factorisation of ${\cal E}$. One could play the same game by substituting $a=3l, 3l\pm 1$ and $b=3m, 3m\pm 1$ in the expression
(\ref{en2}).  After some algebra we find that ${\cal E}=3^2k$ or ${\cal E}=3(3k+1)$ or  ${\cal E}=3k+1$, of course not all the values of $k$ are viable in these expressions. Combining the above results and extracting out the factor of $2$ and $3$, we can write 
${\cal E}=2^{2\alpha}3^{\beta}(6k+1)$. Further factorising $6k+1$ in terms of rational odd primes, we have
 \beq
{\cal E}=2^{2\alpha}3^{\beta}\prod_{i=1}^{I}(6l_i-1)^{2\lambda_i}\prod_{j=1}^{J}(6m_j+1)^{\nu_j}
\label{en4}
\eeq
and which gives the viable energies given by the expression (\ref{en2}) modulo the factor $\frac{8\hbar^2\pi^2}{9\mu L^2}$. 
In (\ref{en4}),  $(6l_i-1)$ and $(6m_j+1)$ are rational odd  primes. Like the previous case we notice again that the $6l_i-1$ type primes always have even powers and $6m_j+1$ type prime can have any power even or odd.

\subsection{Unified formula}
The equations (\ref{en3}) and (\ref{en4}) can be condensed into:
\beq
{\cal E}=(Z-1)^{2\alpha}Z^{\beta}\prod_{i=1}^{I}(2l_iZ-1)^{2\lambda_i}\prod_{j=1}^{J}(2m_jZ+1)^{\nu_j},
\label{enu}
\eeq
where $Z=2$ and $Z=3$ recover the expressions  (\ref{en3}) and (\ref{en4}) respectively.

Before presenting the solution to the degeneracy problem we introduce a notation
which will be used in the solution. We define the function $y(\tau_1,\tau_2,\tau_3,\cdots)$ as,
$$y(\tau_1,\tau_2,\tau_3,\cdots)=\left\{
	\begin{array}{ll}
		1;  & {\rm all} ~ \tau_i={\rm even}  \\
		0; &  {\rm otherwise}. 
	\end{array}\right. $$
For example, $y(0,4,2,6,6,8)=1$,  $y(2,4,3,6,5)=0$ and $y(2,1,4,4)=0$. \\

Degeneracy for both the cases is given by a single expression,
\beq
 {\cal D}^{(2)}_{sq/et}({\cal E})=\prod_{j=1}^{J}(1+\nu_j)-\delta_{1,y(\beta,\nu_1,\nu_2,\nu_3,\cdots,\nu_J)},
\eeq
where the second term is a Kronecker $\delta_{i,j}$ and  we have denoted subscripts `$sq$' for the square box and `$et$' for the equilateral triangular box problems respectively. This expression is independent of $\alpha,\lambda_i,l_i, m_j$ and $Z$. Note that ${\cal D}^{(2)}_{sq}({\cal E})=c_2({\cal E})$.\\

\footnotesize
\noindent {\bf Example:} Let us take ${\cal E}={\bf 6253}$, which is factorised as $6253=13^2.37$.  For the square box problem,  $6253=(4\times 3+1)^{2}(4\times 9+1)$, comparing this with (\ref{en3}), we read  $\beta=0$, $m_1=3$, $\nu_1=2$, $m_2=9$ and  $\nu_2=1$. So, 
${\cal D}^{(2)}_{sq}{({\bf 6253})}=(1+\nu_1)(1+\nu_2)-\delta_{1,y(\beta,\nu_1,\nu_2)}=(1+2)(1+1)-\delta_{1,y(0,2,1)}
=3.2-\delta_{1,0}=6-0={\bf 6}.$\\
The  same energy is also viable for equilateral triangular box, $6253=(6\times 2+1)^{2}(6\times 6+1)$, comparing it with the expression (\ref{en4}), we have $\alpha=0$, $\beta=0$, $m_1=2$, $\nu_1=2$, $m_2=6$ and $\nu_2=1$, 
giving the degeneracy,\\
$ {\cal D}^{(2)}_{et}{({\bf 6253})}=(1+\nu_1)(1+\nu_2)-\delta_{1,y(\beta,\nu_1,\nu_2)}=(1+2)(1+1)-\delta_{1,y(0,2,1)}=3.2-\delta_{1,0}=6-0={\bf 6}.$ In this example the degeneracy is the same for both the problems. In the following we present examples where the degeneracies in these two cases are different.\\

\noindent {\bf Example:} Take, ${\cal E}={\bf 16900}=2^2.5^2.13^2$. For square box,  $16900=2^2(4\times 1+1)^{2}(4\times 3+1)^{2}$, comparing this with (\ref{en3}), we read  $\beta=2$, $m_1=1$, $\nu_1=2$, $m_2=3$ and  $\nu_2=2$. So, 
${\cal D}^{(2)}_{sq}{({\bf 16900})}=(1+\nu_1)(1+\nu_2)-\delta_{1,y(\beta,\nu_1,\nu_2,)}=(1+2)(1+2)-\delta_{1,y(2,2,2)}
=3.3-\delta_{1,1}=9-1={\bf 8}.$\\
For equilateral triangular box, $16900=2^2(6\times 1-1)^{2}(6\times 2+1)^{2}$, comparing it with the expression (\ref{en4}), we have $\alpha=1$, $\beta=0$, $l_1=1$, $\lambda_1=1$, $m_1=2$ and $\nu_1=2$, 
giving the degeneracy,\\
$ {\cal D}^{(2)}_{et}{({\bf 16900})}=(1+\nu_1)-\delta_{1,y(\beta,\nu_1)}=(1+2)-\delta_{1,y(0,2)}=3-\delta_{1,1}=3-1={\bf 2}.$\\

\noindent {\bf Example:}  Take ${\cal E}={\bf 3530}=2.5.353=2(4\times 1+1)(4\times 88+1)$,  $\beta=1$, $m_1=1$, $\nu_1=1$, $m_2=88$ and $\nu_{2}=1$ using  (\ref{en3}).
${\cal D}^{(2)}_{sq}{({\bf 3530})}=(1+\nu_1)(1+\nu_{2})-\delta_{1,y(\beta,\nu_1,\nu_2)}=(1+1)(1+1)-\delta_{1,y(1,1,1)}=4-\delta_{1,0}=4-0={\bf 4}.$\\
Factorisation of 3530 cannot be cast in the form given in (\ref{en4}) (since the power of $5(=6\times 1-1)$ is odd), so this is not a viable energy for an equilateral triangular box and hence ${\cal D}^{(2)}_{et}{({\bf 3530})}=0.$ \\

\noindent {\bf Example:}  ${\cal E}={\bf 2100}=2^2.3.5^2.7=2^2.3.(6\times 1-1)^2(6\times 1+1)$, comparing with  (\ref{en4}),  $\alpha=1$, $\beta=1$, $l_1=1$, $\lambda_1=1$,
$m_1=1$ and  $\nu_{1}=1$.
${\cal D}^{(2)}_{et}{({\bf 2100})}=(1+\nu_1)-\delta_{1,y(\beta,\nu_1)}=(1+1)-\delta_{1,y(1,1)}=2-\delta_{1,0}=2-0={\bf 2}.$\\
Factorisation of 2100 cannot be cast in the form (\ref{en3})(since powers of $3(=4\times 1-1)$ and $7(=4\times 2-1)$ are odd), so this is not a viable energy for a square box and hence ${\cal D}^{(2)}_{sq}{({\bf 2100})}=0.$\\ 

\noindent {\bf Example:}  It could be easily seen that  ${\cal E}={\bf 120}=2^3.3.5$, which is not a viable energy in either of the cases. It gives ${\cal D}^{(2)}_{sq}{({\bf 120})}=0={\cal D}^{(2)}_{et}{({\bf 120})}.$

\small

\section{$\kappa=3$ case: Particle in a cubic box}

So the problem, in its simplest form, is about finding the degeneracy, ${\cal {\cal D}}({\cal E})$ of the energy ${\cal E}$ given by,
\beq
{\cal E}=a^2+b^2+c^2; \qquad a,b,c \in \mathbb{N},
\label{en3d}
\eeq
or in other words {\it how many ways can a positive integer be represented as sum of three natural number squares?}

\subsection{Gauss-Legendre Three Square Theorem: ${\cal E}=4^j(8k+7)$}

The Gauss-Legendre three square theorem \cite{Grosswald_1985} could be proven in the following way by enumerating all the possibilities of $a$, $b$ and $c$ in the expression of ${\cal E}$.
There are the following four possibilities ($i,l,m\in\mathbb{N}$):
\begin{enumerate}
\item{{\it $ a, b, c, $ all even }: $a=2i, ~b=2l$ and $c=2m$,  then ${\cal E}=4(i^2+l^2+m^2)$  which is either 0 or 4 modulo 8 depending on whether $[i^2+l^2+m^2]$ is even or odd. }

\item{{\it $ a, b, c, $ two even and one odd} : $a=2i, ~b=2l$ and $c=2m+1$,  then ${\cal E}=4[i^2+l^2+m(m+1)]+1$  which is either 1 or 5 modulo 8 depending on whether $i^2+l^2$ is even or odd. }

\item{{\it $ a, b, c, $ one even and two odd}:  $a=2i, ~b=2l+1$ and $c=2m+1$,  then ${\cal E}=4[i^2+l(l+1)+m(m+1)]+2$ which is either 2 or 6 modulo 8 depending on whether $i^2$ is even or odd. }

\item{{\it $ a, b, c,$ all odd} : $a=2i+1, ~b=2l+1$ and $c=2m+1$,  then ${\cal E}=4[i(i+1)+l(l+1)+m(m+1)]+3$ which is 3 modulo 8 since products of consecutive integers like $i(i+1)$ are always even and which in turn makes the factor $[i(i+1)+l(l+1)+m(m+1)]$ even. }
\end{enumerate}
Therefore the numbers of the type $8k+7$ are absent in ${\cal E}$, so 
\beq
{\cal {\cal D}}(8k+7)=0.
\label{th1}
\eeq
Now suppose, ${\cal E'}=p'^2+q'^2+r'^2=4{\cal E}$. We see that among the above four possibilities only the case $p',q,'r'$ all even survives since  ${\cal E'}$ is zero modulo 4. Hence one can write $p'=2p,$ $q'=2q$ and $r'=2r$, which then results in $ {\cal E}=p^2+q^2+r^2$, for each set of distinct $p',q,'r'$ satisfying  $p'^2+q'^2+r'^2={\cal E'}$. 
On the other hand for every distinct set $p,q,r,$ satisfying  $ p^2+q^2+r^2={\cal E},$ there would be a set $p'=2p,$ $q'=2q$ and $r'=2r$  satisfying ${\cal E'}=4{\cal E}.$ This leads to the result ${{\cal D}}({\cal E'})={ {\cal D}}(2^2{\cal E})={ {\cal D}}(4{\cal E})={ {\cal D}}({\cal E})$. Applying this result iteratively we arrive at,  
\beq
{ {\cal D}}(2^{2j}{\cal E})={ {\cal D}}(4^j{\cal E})={ {\cal D}}(4^{j-1}{\cal E})={{\cal D}}(4^{j-2}{\cal E})=\cdots={ {\cal D}}(4^2{\cal E})={ {\cal D}}(4{\cal E})={ {\cal D}}({\cal E}).
\label{th2}\eeq
Combining the results (\ref{th1}) and (\ref{th2}) we have the Gauss-Legendre three square theorem \cite{Grosswald_1985} ,
\beq
{\cal {\cal D}}(4^j(8k+7))=0.
\eeq
\subsection{ ${\cal E}=2^j$}
Here we quickly mention the case when ${\cal E}$ has only powers of two, that is ${\cal E}=2^j$. By simple inspection one finds ${{\cal D}}(1)=0$ and ${ {\cal D}}(2)=0$. Using (\ref{th2}) with the former condition, one finds ${ {\cal D}}(2^{2j})={ {\cal D}}(2^{2j}.1)={{\cal D}}(1)=0$ and using  (\ref{th2}) with the latter, one has  ${ {\cal D}}(2^{2j+1})={ {\cal D}}(2^{2j}.2)={{\cal D}}(2)=0.$ Together they result in $ {{\cal D}}(2^{j})=0.$ 

The result (\ref{th2}) simplifies the task considerably. 
We now have to worry about only two types of positive integers: the first type,  where ${\cal E}$ is free from any factors of two and the second type of the form $2{\cal E}$.
\subsection{ ${\cal E}=p^j$ or  ${\cal E}=2p^j$  , where $p$ is a rational odd prime}
We show a systematic procedure of building a formula for the most general case of ${\cal E}$.
We first address the degeneracy counting for the  ${\cal E}=p^j$, where  $p$  is a rational odd prime.
When the power $j$ is odd, our degeneracy formula involves the class number of imaginary quadratic fields.
We denote $h(-d)$ as the class number of imaginary quadratic field $\mathbb{Q}(\sqrt{-d})$, where $d>0$ and is a squarefree integer. Note that $d$ is not the discriminant. A study of the list of degeneracies results in the following.
\subsubsection{$p=4i+1:$}
\begin{eqnarray}
&{\cal E}=p^{2m-1}; \qquad  &{{\cal D}}({\cal E})=\frac{3}{2}\left[h(-p)\frac{(p^m-1)}{(p-1)}-2m\right];\label{prime1}\\
&{\cal E}=p^{2m}; \qquad  &{\cal D}({\cal E})={3}\left[\frac{(p^m-1)}{4}-m\right];\\
&{\cal E}=2p^{2m-1}; \qquad  &{\cal D}({\cal E})=\frac{3}{2}\left[h(-2p)\frac{(p^m-1)}{(p-1)}-2m\right];\\
&{\cal E}=2p^{2m}; \quad & \left\{
	\begin{array}{ll}
		 {\cal D}({\cal E})=\displaystyle{\frac{3}{2}}\left[\displaystyle{\frac{(p+1)}{(p-1)}}(p^m-1)-2m\right]\quad &p=8l-3; \\
	 {\cal D}({\cal E})=3\left[\displaystyle{\frac{(p^m-1)}{2}}-m\right] \quad & p=8l+1.
	\end{array}\right.\label{prime2}
\end{eqnarray}
\footnotesize
{\bf Example:}   ${\cal E}=13^3={\bf 2197}$; $m=2$ and $h(-13)=2$\footnote{ \texttt{http://www.numbertheory.org/php/classnonegtable\_qfields.html}}; 
 ${\cal D}({\bf 2197})=\displaystyle{\frac{3}{2}}\left[2\displaystyle{\frac{(13^2-1)}{(13-1)}}-2.2\right]={\bf 36}.$

\noindent {\bf Example:}   ${\cal E}=2.29^4={\bf 1414562};$  $m=2$ and $p=29=8.4-3;$ Use first of formula (\ref{prime2}) to get  ${\cal D}({\bf 1414562})=\displaystyle{\frac{3}{2}}\left[\displaystyle{\frac{(29+1)}{(29-1)}}(29^2-1)-2.2\right]=\displaystyle{\frac{3}{2}}\left[(29+1)^2-2.2\right]={\bf 1344}.$
\normalsize
\subsubsection{$p=4i-1:$}
\begin{eqnarray}
&{\cal E}=p^{2m-1}; \quad &\left\{
	\begin{array}{ll}
		  {\cal D}({\cal E})=0 \quad &p=8l-1; \\
		 {\cal D}({\cal E})=3h(-p)\displaystyle{\frac{(p^m-1)}{(p-1)}} \quad &p=8l+3;
	\end{array}\right.\\
& {\rm Except~for } ~p=3; \quad {\cal E}=3^{2m-1}; \quad & {\cal D}({\cal E})=\displaystyle{\frac{(3^m-1)}{2}};\\
&{\cal E}=p^{2m}; \quad  &{\cal D}({\cal E})=\frac{3}{4}\frac{(p+1)}{(p-1)}(p^m-1);\label{prime3}\\
&{\cal E}=2p^{2m-1}; \quad  &{\cal D}({\cal E})=\frac{3}{2}h(-2p)\frac{(p^m-1)}{(p-1)}; \label{prime4} \\
&{\cal E}=2p^{2m}; \quad &\left\{
	\begin{array}{ll}
		 {\cal D}({\cal E})=\displaystyle{\frac{3}{2}}\displaystyle{\frac{(p+1)}{(p-1)}}(p^m-1)\quad &p=8l-1; \\
		{\cal D}({\cal E})=\displaystyle{\frac{3}{2}{(p^m-1)}}\quad & p=8l+3; 
	\end{array}\right.
\end{eqnarray}
\footnotesize
{\bf Example:}   ${\cal E}=11^8={\bf 214358881}$; $m=4,$  Use formula (\ref{prime3}) to get \\
 ${\cal D}({\bf 214358881})=\displaystyle{\frac{3}{4}\frac{(11+1)}{(11-1)}(11^4-1)}={\bf 13176}.$

\noindent {\bf Example:}   ${\cal E}=2.23^5={\bf 12872686};$ $m=3$ and $h(-2.23)=h(-46)=4$;  Use formula (\ref{prime4}) to get \\
 ${\cal D}({\bf 12872686})=\displaystyle{\frac{3}{2}.4\frac{(23^3-1)}{(23-1)}}={\bf 3318}.$

\small

\subsection{Form ${\cal E}$ and $2{\cal E}$, where ${\cal E}$ is factorised into even powers of \\
rational odd primes}

\subsubsection{ Case ${\cal E}$}
In the following $p_i$ and $q_j$ are rational odd primes.
$p_i=4m_i-1$ and $q_j=4n_j+1$; $i,j,m_i,n_j \in \mathbb{N}$; $w_i,x_j \in \mathbb{Z}_{\geq 0}$.
$${\cal E}=\prod_{i=1}^{M}p_i^{2w_i}\prod_{j=1}^{N}q_j^{2x_j};$$  
\beq
 {\cal D}({\cal E})=\frac{3}{4}\left[\prod_{i=1}^{M}\left\{\displaystyle{\frac{(p_i+1)}{(p_i-1)}}(p_i^{w_i}-1)+1 \right\}\prod_{j=1}^{N}q_j^{x_j}-2\prod_{j=1}^{N}(2x_j+1)+1\right]\label{prime5}
\eeq
\footnotesize
{\bf Example:}   ${\cal E}=3^6.5^4.11^2={\bf 55130625};$  Here, $M=2$ and $N=1$; $p_1=3,p_2=11,q_1=5$ and $w_1=3,w_2=1,x_1=2$. Substituting in (\ref{prime5}), we have,\\
$ {\cal D}({\bf 55130625})=\displaystyle{\frac{3}{4}}\left[\left\{\displaystyle{\frac{(3+1)}{(3-1)}}(3^3-1)+1\right\}\left\{\displaystyle{\frac{(11+1)}{(11-1)}}(11^1-1)+1\right\}5^2-2(2.2+1)+1\right]$\\
$\phantom{{\cal D}({\bf 55130625})}=\displaystyle{\frac{3}{4}}\left[\left\{52+1\right\}\left\{12+1\right\}25-10+1\right]=\displaystyle{\frac{3}{4}}\left[17225-10+1\right]=\displaystyle{\frac{3}{4}}\left[17216\right]={\bf 12912}.$

\small

\subsubsection{ Case  $2{\cal E}$}
\noindent For evaluating the degeneracies for energies of the type $2{\cal E}$ where ${\cal E}$ factorises into even powers of rational odd primes,
we need to distinguish  the primes modulo 8.  Below $p_i,q_j, r_k$ and $s_l$ are rational odd primes., where $p_i=(8m_i-3)$,~$q_j=(8n_j-1)$, ~$r_k=(8u_k+1)$ and $s_l=(8v_l+3)$;\\
$i,j,k,l,m_i,n_j, u_k\in \mathbb{N}$; $v_l,w_i,x_j,y_k,z_l \in \mathbb{Z}_{\geq 0}$.
$${\cal E'}=2\prod_{i=1}^{M}p_i^{2w_i}\prod_{j=1}^{N}q_j^{2x_j}\prod_{k=1}^{U}r_k^{2y_k}\prod_{l=1}^{V}s_l^{2z_l};$$ 
For the above energy we have the degeneracy,
$${\cal D}({\cal E'})=\frac{3}{2}\left[\prod_{i=1}^{M}\left\{\displaystyle{\frac{(p_i+1)}{(p_i-1)}}(p_i^{w_i}-1)+1 \right\}\prod_{j=1}^{N}\left\{\displaystyle{\frac{(q_j+1)}{(q_j-1)}}(q_j^{x_j}-1)+1 \right\}\prod_{k=1}^{U}r_k^{y_k}\prod_{l=1}^{V}s_l^{z_l}\right.$$
\beq
\left.-\prod_{i=1}^{M}(2w_i+1)\prod_{k=1}^{U}(2y_k+1)\right]
\label{prime6}
\eeq
\footnotesize
{\bf Example:}   ${\cal E'}=2.5^4.7^2.11^2.17^2={\bf 2141851250};$ We read,  $M=1, N=1,U=1$ and $V=1$; $p_1=5,q_1=7,r_1=17, s_1=11$ and $w_1=2,x_1=1,y_1=1,z_1=1$.\\
$ {\cal D}({\bf  2141851250})=\displaystyle{\frac{3}{2}}\left[\left\{\displaystyle{\frac{(5+1)}{(5-1)}}(5^2-1)+1\right\}\left\{\displaystyle{\frac{(7+1)}{(7-1)}}(7^1-1)+1\right\}11^1.17^1-(2.2+1)(2.1+1)\right]$\\
$\phantom{{\cal D}({\bf 2141851250})}=\displaystyle{\frac{3}{2}}\left[\left\{36+1\right\}\left\{8+1\right\}187-5.3\right]=\displaystyle{\frac{3}{2}}\left[62271-15\right]=\displaystyle{\frac{3}{2}}\left[62256\right]={\bf 93384}.$

\small

\subsection{Form ${\cal E}$ and $2{\cal E}$, where ${\cal E}$ is factorised into odd powers of rational odd primes}

\subsubsection{ Case  ${\cal E}$}
In the following $p_i$ and $q_j$ are rational odd primes.
$p_i=4m_i-1$ and $q_j=4n_j+1$;\\ $i,j,m_i,n_j,d,d'\in \mathbb{N}$; $w_i,x_j ,f,g\in \mathbb{Z}_{\geq 0}$.

$${\cal E}=\prod_{i=1}^{M}p_i^{2w_i+1}\prod_{j=1}^{N}q_j^{2x_j+1};\qquad {\rm and} ~ \text{\lq}d\text{\rq}~{\rm is~ defined~as,}~d= \prod_{i=1}^{M}p_i\prod_{j=1}^{N}q_j$$ 

$ i)~M=$ odd; ~ $d=4f+3$;
\beq
 \left\{
	\begin{array}{ll}
		  d=8g+7;  &\quad {\cal D}({\cal E})=0; \\
		  d=8g+3; & \quad {\cal D}({\cal E})=3h(-d)\displaystyle{\prod_{i=1}^{M}\frac{(p_i^{w_i+1}-1)}{(p_i-1)}\prod_{j=1}^{N}\frac{(q_j^{x_j+1}-1)}{(q_j-1)}}
	\end{array}\right.
\label{prime7}
\eeq

$ ii)~M=$ even; ~ $d=4f+1$;
\beq
{\cal D}({\cal E})=\frac{3}{2}\left[h(-d)\displaystyle{\prod_{i=1}^{M}\frac{(p_i^{w_i+1}-1)}{(p_i-1)}\prod_{j=1}^{N}\frac{(q_j^{x_j+1}-1)}{(q_j-1)}}-\delta_{0,M}\prod_{j=1}^{N}2(x_j+1)\right]
\label{prime8}
\eeq
\footnotesize
{\bf Example:}   ${\cal E}=3^5.5^3.17.29={\bf 14974875};$  We read,  $M=1,$ $N=3$; $p_1=3,q_1=5,q_2=17, q_3=29$ and $w_1=2,x_1=1,x_2=0,x_3=0$; $d=3.5.17.29=7395=8.924+3$, we use (\ref{prime7}),\\
$ {\cal D}({\bf 14974875})=3h(-7395)\displaystyle{\frac{(3^{2+1}-1)}{(3-1)}\frac{(5^{1+1}-1)}{(5-1)}\frac{(17^{0+1}-1)}{(17-1)}\frac{(29^{0+1}-1)}{(29-1)}}=3.16.13.6.1.1={\bf 3744},$

\noindent where we have used $h(-7395)=16.$\\
{\bf Example:}   ${\cal E}=5^5.13^3.37={\bf 254028125};$  We read,  $M=0,$ $N=3$; $q_1=5,q_2=13, q_3=37$ and $x_1=2,x_2=1,x_3=0$; $d=5.13.37=2405=4.601+1$, we use (\ref{prime8}),\\
$ {\cal D}({\bf 254028125})=\displaystyle{\frac{3}{2}}\left[h(-2405)\displaystyle{\frac{(5^{2+1}-1)}{(5-1)}\frac{(13^{1+1}-1)}{(13-1)}\frac{(37^{0+1}-1)}{(37-1)}}-\delta_{0,0}2(2+1).2(1+1).2(0+1)\right]$\\
$\phantom{{\cal D}({\bf 254028125})}=\displaystyle{\frac{3}{2}}\left[40.31.14.1-1.6.4.2\right]=\displaystyle{\frac{3}{2}}\left[17360-48\right]=\displaystyle{\frac{3}{2}}\left[17312\right]={\bf 25968},$

\noindent where we have used $h(-2405)=40.$ 

\small

\subsubsection{ Case  $2{\cal E}$}
For the case $2{\cal E}$ we proceed with the following,
$${\cal E'}=2\prod_{i=1}^{M}p_i^{2w_i+1}\prod_{j=1}^{N}q_j^{2x_j+1};\qquad {d'}= 2\prod_{i=1}^{M}p_i\prod_{j=1}^{N}q_j$$ 
\beq
{\cal D}({\cal E'})=\frac{3}{2}\left[h(-d')\displaystyle{\prod_{i=1}^{M}\frac{(p_i^{w_i+1}-1)}{(p_i-1)}\prod_{j=1}^{N}\frac{(q_j^{x_j+1}-1)}{(q_j-1)}}-\delta_{0,M}\prod_{j=1}^{N}2(x_j+1)\right]
\label{prime9}
\eeq
\footnotesize
{\bf Example:}   ${\cal E'}=2.3^7.5^3.7^3.13={\bf 2437958250};$  We read,  $M=2,$ $N=2$; $p_1=3,p_2=7,q_1=5,q_2=13$ and $w_1=3,w_2=1,x_1=1,x_2=0$; $d'=2.3.5.7.13=2730$, we use (\ref{prime9}),\\
$ {\cal D}({\bf 2437958250})=\displaystyle{\frac{3}{2}}\left[h(-2730)\displaystyle{\frac{(3^{3+1}-1)}{(3-1)}\frac{(5^{1+1}-1)}{(5-1)}\frac{(7^{1+1}-1)}{(7-1)}\frac{(13^{0+1}-1)}{(13-1)}}-\delta_{0,2}2(1+1).2(0+1)\right]$\\
$\phantom{{\cal D}({\bf 2437958250})}=\displaystyle{\frac{3}{2}}\left[32.40.6.8.1-0.6.4.2\right]=\displaystyle{\frac{3}{2}}\left[61440\right]={\bf 92160},$

\noindent where we have used $h(-2730)=32.$ 

\small

\subsection{General case of  ${\cal E}$ and $2{\cal E}$, where ${\cal E}$ is factorised into odd as well as even powers of rational odd primes}
\subsubsection{ Case  ${\cal E}$}
Below $p_i,q_j, r_k$ and $s_l$ are rational odd primes where $p_i=(4m_i-1)$,~$q_j=(4n_j+1)$, ~$r_k=(4u_k-1)$ and $s_l=(4v_l+1).$ Further $p_i\neq r_k$($m_i\neq u_k$) and $q_j\neq s_l$($n_j\neq v_l$);
$i,j,k,l,m_i,n_j, u_k,v_l,d,d'\in \mathbb{N}$; $w_i,x_j,y_k,z_l ,f,g\in \mathbb{Z}_{\geq 0}$.
$${\cal E}=\prod_{i=1}^{M}p_i^{2w_i+1}\prod_{j=1}^{N}q_j^{2x_j+1}\prod_{k=1}^{U}r_k^{2y_k}\prod_{l=1}^{V}s_l^{2z_l};\qquad d= \prod_{i=1}^{M}p_i\prod_{j=1}^{N}q_j$$ 
In the above the rational primes $r_j$ and $s_k$ (which have even powers) are now required to be sorted as `Inert'(denoted by $I_a$)  or `Split'(denoted by $S_b$)  in imaginary quadratic field,  $\mathbb{Q}(\sqrt{-d})$. ${\cal E}$ is now factored as the following,
$${\cal E}=\prod_{i=1}^{M}p_i^{2w_i+1}\prod_{j=1}^{N}q_j^{2x_j+1}\prod_{a=1}^{A}I_a^{2c_a}\prod_{b=1}^{B}S_b^{2t_b};\qquad U+V=A+B$$ 
where each of $I_a$ and $S_b$ is either $r_k$ or $s_l$  and each of $c_a$ and $t_b$ is either $y_k$ or $z_l$.  

$i)~M={\rm odd}; \quad  d=4f+3;$
\beq
\left\{
	\begin{array}{ll}
		  d=8g+7;  &{\cal D}({\cal E})=0; \\
		  d=8g+3; & {\cal D}({\cal E})=3h(-d)\displaystyle{\prod_{i=1}^{M}\frac{(p_i^{w_i+1}-1)}{(p_i-1)}\prod_{j=1}^{N}\frac{(q_j^{x_j+1}-1)}{(q_j-1)}\prod_{a=1}^{A}\left\{\displaystyle{\frac{(I_a+1)}{(I_a-1)}}(I_a^{c_a}-1)+1 \right\}}\prod_{b=1}^{B}S_b^{t_b}
	\end{array}\right.
\label{prime10}
\eeq
\noindent $ii)~M={\rm even}; \quad d=4f+1;$
$$\hspace*{-1cm} {\cal D}({\cal E})=\frac{3}{2}\left[h(-d)\displaystyle{\prod_{i=1}^{M}\frac{(p_i^{w_i+1}-1)}{(p_i-1)}\prod_{j=1}^{N}\frac{(q_j^{x_j+1}-1)}{(q_j-1)}\prod_{a=1}^{A}
\left\{\displaystyle{\frac{(I_a+1)}{(I_a-1)}}(I_a^{c_a}-1)+1 \right\}}\prod_{b=1}^{B}S_b^{t_b}\right.$$
\beq
\left.-\delta_{0,M}\prod_{j=1}^{N}2(x_j+1)\prod_{l=1}^{V}(2z_l+1)\right]
\label{prime11}
\eeq
To determine whether a rational odd prime, $p$, is `Inert' type (that is $p$ cannot be factorised in terms of prime ideals) or  `Split' type ( when $p$ can be factorised in terms of prime ideals) in  $\mathbb{Q}(\sqrt{-d})$ one has to calculate the Legendre symbol $\left(\displaystyle{\frac{-d}{p}}\right)$. Now the Legendre symbol,\footnote{$p$ never divides $-d$ exactly since $d$ is a product of the primes $p_i$ and $q_j$ but the prime $p$ is either of the type $r_k$ or $s_l$ (which have only even powers in the factorisation of ${\cal E}$) and as mentioned above they are distinct from $p_i$ and $q_j$, so the ramified primes are absent in this sorting.}
$$\left(\displaystyle{\frac{-d}{p}}\right)=\left\{
	\begin{array}{ll}
		  1;  &\quad {\rm if~ there ~is ~a ~quadratic~ residue~ then}~ p~{\rm is}~{\rm Split} ; \\
		 -1; & \quad  {\rm if~ there ~is ~no ~quadratic~ residue~ then}~ p~{\rm is}~{\rm Inert} .
	\end{array}\right.$$
\footnotesize
{\bf Example:}  ${\cal E}=3^3.5^3.7.11.13^2.19^2={\bf 15854713875};$   We read,  $M=3,$ $N=1,$ $U=1$ and $V=1$;\\ $p_1=3,p_2=7,p_3=11,q_1=5,r_1=19,s_1=13$ and
 $w_1=1,w_2=0,w_3=0,x_1=1,y_1=1,z_1=1;$ $d=3.7.11.5={\bf 1155}$ and $ h(-1155)=8;$
 We now sort `Inert' and `Split' primes in  $\mathbb{Q}(\sqrt{-1155})$. \\
$-d=-1155$. We calculate Legendre symbol,
for $r_1=19;$\\
$\left(\displaystyle{\frac{-1155}{19}}\right)=\left(\displaystyle{\frac{-19.60-15}{19}}\right)
=\left(\displaystyle{\frac{-15}{19}}\right)=4=2^2\equiv 1;$ quadratic residue $\Rightarrow r_1=19=S_1\equiv$ {\bf Split} and $t_1=z_1=1$.\\
For $s_1=13;$ $\left(\displaystyle{\frac{-1155}{13}}\right)=\left(\displaystyle{\frac{-13.80-11}{13}}\right)
=\left(\displaystyle{\frac{-11}{13}}\right)=2\equiv -1;$ not a quadratic residue $\Rightarrow s_1=13=I_1\equiv$ {\bf Inert} and  $c_1=y_1=1$. We use formula (\ref{prime10}) since
$d=8.144+3$.\\
 ${\cal D}({\bf 15854713875})=3h(-1155)\displaystyle{\frac{(3^{1+1}-1)}{(3-1)}\frac{(7^{0+1}-1)}{(7-1)}\frac{(11^{0+1}-1)}{(11-1)}\frac{(5^{1+1}-1)}{(5-1)}\left\{\displaystyle{\frac{(13+1)}{(13-1)}}(13^{1}-1)+1 \right\}}19^{1}$\\
 $\phantom{{\cal D}({\bf 15854713875})}=3.8.\displaystyle{\frac{8}{2}.1.1\frac{24}{4}.\left\{14+1 \right\}}.19$=3.8.4.6.15.19={\bf 164160}.\\
{\bf Example:}  ${\cal E}=3^2.5^3.7^4.13.17^2={\bf 10148126625};$   We read,  $M=0,$ $N=2,$ $U=2$ and $V=1$;\\ $q_1=5,q_2=13,r_1=3,r_2=7,s_1=17$ and
 $x_1=1,x_2=0,y_1=1,y_2=2,z_1=1;$\\ $d=5.13={\bf 65};$ and   $h(-65)=8.$
 We now sort `Inert' and `Split' primes in  $\mathbb{Q}(\sqrt{-65})$. \\
$-d=-65$. We calculate Legendre symbol,
for $r_1=3;$\\
$\left(\displaystyle{\frac{-65}{3}}\right)=\left(\displaystyle{\frac{-3.21-2}{3}}\right)
=\left(\displaystyle{\frac{-2}{3}}\right)=1;$ quadratic residue $\Rightarrow r_1=3=S_1\equiv$ {\bf Split} and $t_1=y_1=1$.\\
For $r_2=7;$ $\left(\displaystyle{\frac{-65}{7}}\right)=\left(\displaystyle{\frac{-7.9-2}{7}}\right)
=\left(\displaystyle{\frac{-2}{7}}\right)=5\equiv -1;$ not a quadratic residue $\Rightarrow r_2=7=I_1\equiv$ {\bf Inert} and  $c_1=y_2=2$.\\
For $s_1=17;$ $\left(\displaystyle{\frac{-65}{17}}\right)=\left(\displaystyle{\frac{-17.3-14}{17}}\right)
=\left(\displaystyle{\frac{-14}{17}}\right)=3\equiv -1;$ not a quadratic residue $\Rightarrow s_1=17=I_2\equiv$ {\bf Inert} and  $c_2=z_1=1$. We use formula (\ref{prime11}) since $d=4.16+1$.\\
 ${\cal D}({\bf 10148126625})=\displaystyle{\frac{3}{2}}\left[h(-65)\displaystyle{\frac{(5^{1+1}-1)}{(5-1)}\frac{(13^{0+1}-1)}{(13-1)}
\left\{\frac{(7+1)}{(7-1)}(7^{2}-1)+1 \right\}\left\{\frac{(17+1)}{(17-1)}(17^{1}-1)+1 \right\}}3^{1}\right.$\\
$-\delta_{0,0}2(1+1).2(0+1).(2.1+1)\bigg]=\displaystyle{\frac{3}{2}}\left[8.\displaystyle{\frac{24}{4}.1\left\{64+1 \right\}\left\{18+1 \right\}}.3-4.2.3\right]=3\left[88920-12\right]={\bf 266724}.$\\

\small

\subsubsection{ Case $2{\cal E}$}
$${\cal E'}=2\prod_{i=1}^{M}p_i^{2w_i+1}\prod_{j=1}^{N}q_j^{2x_j+1}\prod_{k=1}^{U}r_k^{2y_k}\prod_{l=1}^{V}s_l^{2z_l};\qquad d'=2 \prod_{i=1}^{M}p_i\prod_{j=1}^{N}q_j$$ 
$${\cal E'}=2\prod_{i=1}^{M}p_i^{2w_i+1}\prod_{j=1}^{N}q_j^{2x_j+1}\prod_{a=1}^{A}I_a^{2c_a}\prod_{b=1}^{B}S_b^{2t_b};\qquad U+V=A+B$$ 
Like the earlier case, in the above $r_k$ and $s_l$ are now sorted as `Inerts', $I_a,$ and `Splits', $S_b,$ in $\mathbb{Q}(\sqrt{-{d'}}).$  
$$\hspace*{-1cm} {\cal D}({\cal E'})=\frac{3}{2}\left[h(-d')\displaystyle{\prod_{i=1}^{M}\frac{(p_i^{w_i+1}-1)}{(p_i-1)}\prod_{j=1}^{N}\frac{(q_j^{x_j+1}-1)}{(q_j-1)}\prod_{a=1}^{A}
\left\{\displaystyle{\frac{(I_a+1)}{(I_a-1)}}(I_a^{c_a}-1)+1 \right\}}\prod_{b=1}^{B}S_b^{t_b}\right.$$
\beq
\left.-\delta_{0,M}\prod_{j=1}^{N}2(x_j+1)\prod_{l=1}^{V}(2z_l+1)\right]
\label{prime12}
\eeq
\footnotesize
{\bf Example:}  ${\cal E}=2.3^5.5^4.7^3.13^2={\bf 17607476250};$   We read,  $M=2,$ $N=0,$ $U=0$ and $V=2$;\\ $p_1=3,p_2=7,s_1=5,s_2=13$ and
 $w_1=2,w_2=1,z_1=2,z_2=1;$ $d'=2.3.7={\bf 42}$ and $ h(-42)=4;$
 We now sort `Inert' and `Split' primes in  $\mathbb{Q}(\sqrt{-42})$. \\
$-d=-42$. Next we calculate Legendre symbol,
for $s_1=5;$\\
$\left(\displaystyle{\frac{-42}{5}}\right)=\left(\displaystyle{\frac{-5.8-2}{5}}\right)
=\left(\displaystyle{\frac{-2}{5}}\right)=3\equiv -1;$ not a quadratic residue $\Rightarrow s_1=5=I_1\equiv$ {\bf Inert} and $c_1=z_1=2$. For $s_2=13;$\\
$\left(\displaystyle{\frac{-42}{13}}\right)=\left(\displaystyle{\frac{-13.3-3}{13}}\right)
=\left(\displaystyle{\frac{-3}{13}}\right)=10\equiv 1;$ a quadratic residue $\Rightarrow s_2=13=S_1\equiv$ {\bf Split} and  $t_1=z_2=1$. We use formula (\ref{prime12})\footnote{ The second term (containing a Kronecker delta) of the expression (\ref{prime12}) vanishes in this case since $M=2.$},\\
 ${\cal D}({\bf 15854713875})=\displaystyle{\frac{3}{2}}\left[h(-42)\displaystyle{\frac{(3^{2+1}-1)}{(3-1)}\frac{(7^{1+1}-1)}{(7-1)}\left\{\displaystyle{\frac{(5+1)}{(5-1)}}(5^{2}-1)+1 \right\}}13^{1}\right]$\\
 $\phantom{{\cal D}({\bf 15854713875})}=\displaystyle{\frac{3}{2}}\left[4\displaystyle{\frac{26}{2}8\left\{36+1\right\}}13\right]=3.2.13.8.37.13={\bf 300144}.$\\

\small

To sum up we have presented the degeneracy counting of states available for a particle trapped in a cubic box.  This is a classic problem of number theory
and this problem was attacked in several ways in the last two hundred years.  We believe our results are new and the solution is presented in this particular
form for the first time.  We have devised the formulae by studying the list of degeneracies. We do not have any concrete proofs for these results,
so they could be treated as conjectures till they are proven right or wrong.
We have tried to make the note self-contained. For calculations, the only input required from the outside  is the class number, $h(-d)$, of   imaginary quadratic field $\mathbb{Q}(\sqrt{-d})$
which is readily available from various lists available over the web. The class number for smaller $-d$ can also be calculated by the method shown in the book \cite{Lemmermeyer_2021}.
The rest can be computed easily as shown in the explicit examples. Some of the special cases of the above make for handy formulae, we present them in the appendix A. Appendix B gives the list of degeneracies for the first five hundred numbers.

\section*{Acknowledgements}
The author thanks Rohan Raha and Tanmoy Ghosh, with whom discussions on the degeneracy problem in two dimensions  were initiated. Thanks are also due to Soumangsu Bhusan Chakraborty
 for improving the author's understanding of the imaginary quadratic fields.

\section*{Appendix A}

In the following  $j, k,l, n_j, w_i, x_j \in  \mathbb{Z}_{0}^+,  m, m_i\in  \mathbb{N}$
\begin{enumerate}
\item{~$p=$prime$=6l+1$;
\beq
{\cal E}=3^{2m-1}p^{2j}; \qquad {\cal D}({\cal E})=\displaystyle{\frac{(3^m-1)}{2}{p^j}}
\eeq}
\item{$~q=$prime$=6l+5$;
\beq
{\cal E}=3^{2m-1}q^{2j}; \qquad   {\cal D}({\cal E})=\displaystyle{\frac{(3^m-1)}{2}}\left[\displaystyle{\frac{(q+1)}{(q-1)}}(q^j-1)+1\right]
\eeq}
\item{ \beq
{\cal E}=3^{2j}; \qquad  {\cal D}({\cal E})=\displaystyle{\frac{3}{2}{(3^j-1)}}
\eeq}
\item{
\beq
{\cal E}=3p^{2m-1}; ~\left\{
	\begin{array}{ll}
		p={\rm prime}=4l+3;  &\qquad {\cal D}({\cal E})=\displaystyle{\frac{3}{2}}h(-3p)\displaystyle{\frac{(p^m-1)}{(p-1)}}\\
		p={\rm prime}=8l+1; & \qquad  {\cal D}({\cal E})=3h(-3p)\displaystyle{\frac{(p^m-1)}{(p-1)}}
	\end{array}\right.
\eeq}
\item{
\beq
{\cal E}=3p^{2j}q^{2l}; \qquad {\cal D}({\cal E})= {\cal D}({3p^{2j}}){\cal D}({3q^{2l}});\quad p,q={\rm rational~odd~ prime}
\eeq}

\item{ ~$p_i=$prime$=(6m_i-1)$ and $q_j=$prime$=(6n_j+1),$ 
\beq
{\cal E}=3^{2m-1}\prod_{i=1}^{M}p_i^{2w_i}\prod_{j=1}^{N}q_j^{2x_j};~~ {\cal D} ({\cal E})=\frac{(3^m-1)}{2}\prod_{i=1}^{M}\left\{\displaystyle{\frac{(p_i+1)}{(p_i-1)}}(p_i^{w_i}-1)+1 \right\}\prod_{j=1}^{N}q_j^{x_j}
\eeq}

\item{
\beq
{\cal E}=4k+3; \qquad \left\{
	\begin{array}{ll}
		{\cal E}=8l+7;  &  \qquad {\cal D}({\cal E})=0 \\
		{\cal E}=8l+3~ ({\rm squarefree});& \qquad {\cal D}({\cal E})=3h(-{\cal E})
	\end{array}\right.
\eeq}
\qquad\qquad Except ${\cal E}=3~(l=0); \qquad {\cal D}(3)=1$
\item{~$p_i={\rm prime}={4m_i-1}$ and  $q_j={\rm prime}={4n_j+1}$
\beq
{\cal E}=4k+1; \qquad {\cal E}=\prod_{i=1}^{2M}p_i\prod_{j=1}^{N}q_j;   \qquad {\cal D}({\cal E})=\frac{3}{2}\left[h(-{\cal E})-\delta_{0,M}2^N\right]
\eeq}

\end{enumerate}

\section*{Appendix B}

List of degeneracy of  a cubic box up to ${\cal E}=500$ generated numerically by Mathematica\textsuperscript{\textregistered}, in $\{ {\cal E}, {\cal D}{(\cal E)}\}$ format. For the absent integers, the corresponding degeneracy is zero, that is, these integers cannot be represented as sum of three natural number squares.

\bigskip

\footnotesize

\noindent\texttt{\{3, 1\}, \{6, 3\}, \{9, 3\}, \{11, 3\}, \{12, 1\}, \{14, 6\}, \{17, 3\}, \{18, 3\}, \{19, 3\}, \{21, 6\},
\{22, 3\}, \{24, 3\}, \{26, 6\}, \{27, 4\}, \{29, 6\}, \{30, 6\}, \{33, 6\}, \{34, 3\}, \{35, 6\}, \{36, 3\},
\{38, 9\},\\
\{41, 9\}, \{42, 6\}, \{43, 3\}, \{44, 3\}, \{45, 6\}, \{46, 6\}, \{48, 1\}, \{49, 6\}, \{50, 6\},
\{51, 6\},\\
\{53, 6\}, \{54, 12\}, \{56, 6\}, \{57, 6\}, \{59, 9\}, \{61, 6\}, \{62, 12\}, \{65, 6\},
\{66, 12\}, \{67, 3\}, \{68, 3\}, \{69, 12\}, \{70, 6\}, \{72, 3\}, \{73, 3\}, \{74, 12\}, \{75, 7\},
\{76, 3\}, \{77, 12\}, \{78, 6\}, \{81, 12\}, \{82, 3\}, \{83, 9\}, \{84, 6\}, \{86, 15\}, \{88, 3\},
\{89, 15\}, \{90, 12\}, \{91, 6\}, \{93, 6\}, \{94, 12\}, \{96, 3\}, \{97, 3\}, \{98, 12\}, \{99, 9\},
\{101, 18\}, \{102, 6\}, \{104, 6\}, \{105, 12\},\\
\{106, 6\}, \{107, 9\}, \{108, 4\}, \{109, 6\},
\{110, 18\}, \{113, 9\}, \{114, 12\}, \{115, 6\}, \{116, 6\},\\
\{117, 12\}, \{118, 9\}, \{120, 6\},
\{121, 9\}, \{122, 12\}, \{123, 6\}, \{125, 12\}, \{126, 18\}, \{129, 18\}, \{131, 15\}, \{132, 6\},
\{133, 6\}, \{134, 21\}, \{136, 3\}, \{137, 9\}, \{138, 12\}, \{139, 9\}, \{140, 6\}, \{141, 12\},
\{142, 6\}, \{144, 3\}, \{145, 6\}, \{146, 21\}, \{147, 7\}, \{149, 18\}, \{150, 15\}, \{152, 9\},
\{153, 15\}, \{154, 12\}, \{155, 12\}, \{157, 6\}, \{158, 12\}, \{161, 24\}, \{162, 12\}, \{163, 3\},
\{164, 9\}, \{165, 12\}, \{166, 15\}, \{168, 6\}, \{169, 6\}, \{170, 12\}, \{171, 15\}, \{172, 3\},
\{173, 18\}, \{174, 18\}, \{176, 3\}, \{177, 6\}, \{178, 9\}, \{179, 15\}, \{180, 6\}, \{181, 12\},
\{182, 18\}, \{184, 6\}, \{185, 18\}, \{186, 18\}, \{187, 6\}, \{189, 24\}, \{190, 6\}, \{192, 1\},
\{193, 3\}, \{194, 27\}, \{195, 12\}, \{196, 6\}, \{197, 12\}, \{198, 15\}, \{200, 6\}, \{201, 18\},
\{202, 6\}, \{203, 12\}, \{204, 6\}, \{205, 6\}, \{206, 30\}, \{209, 30\}, \{210, 12\}, \{211, 9\},
\{212, 6\}, \{213, 12\}, \{214, 9\}, \{216, 12\}, \{217, 12\}, \{218, 12\}, \{219, 12\}, \{221, 18\},
\{222, 18\}, \{224, 6\}, \{225, 15\}, \{226, 9\}, \{227, 15\}, \{228, 6\}, \{229, 12\}, \{230, 30\},
\{233, 15\}, \{234, 24\}, \{235, 6\}, \{236, 9\}, \{237, 18\}, \{238, 12\}, \{241, 15\}, \{242, 15\},
\{243, 13\}, \{244, 6\}, \{245, 18\}, \{246, 18\}, \{248, 12\}, \{249, 18\}, \{250, 12\}, \{251, 21\},\\
\{253, 6\}, \{254, 24\}, \{257, 21\}, \{258, 12\}, \{259, 12\}, \{260, 6\}, \{261, 24\}, \{262, 9\},
\{264, 12\}, \{265, 6\}, \{266, 30\}, \{267, 6\}, \{268, 3\}, \{269, 30\}, \{270, 24\}, \{272, 3\},
\{273, 12\}, \{274, 15\}, \{275, 15\}, \{276, 12\}, \{277, 6\}, \{278, 21\}, \{280, 6\}, \{281, 27\},
\{282, 12\}, \{283, 9\}, \{285, 24\}, \{286, 18\}, \{288, 3\}, \{289, 9\}, \{290, 24\}, \{291, 12\},
\{292, 3\}, \{293, 24\}, \{294, 21\}, \{296, 12\}, \{297, 24\}, \{298, 6\}, \{299, 24\}, \{300, 7\},
\{301, 12\}, \{302, 18\}, \{304, 3\}, \{305, 18\}, \{306, 27\}, \{307, 9\}, \{308, 12\}, \{309, 18\},
\{310, 12\}, \{312, 6\}, \{313, 9\}, \{314, 36\}, \{315, 18\}, \{317, 12\}, \{318, 18\}, \{321, 30\},
\{322, 12\}, \{323, 12\}, \{324, 12\}, \{325, 12\}, \{326, 33\}, \{328, 3\},\\ 
\{329, 36\}, \{330, 12\}, \{331, 9\}, \{332, 9\}, \{333, 12\}, \{334, 18\}, \{336, 6\}, \{337, 9\}, \{338, 18\}, \{339, 18\},
\{341, 42\}, \{342, 27\}, \{344, 15\}, \{345, 12\}, \{346, 12\}, \{347, 15\}, \{349, 18\},\\
\{350, 30\}, \{352, 3\}, \{353, 21\}, \{354, 24\}, \{355, 12\}, \{356, 15\}, \{357, 12\}, \{358, 9\}, \{360, 12\},
\{361, 15\}, \{362, 24\}, \{363, 13\}, \{364, 6\}, \{365, 24\}, \{366, 18\}, \{369, 33\}, \{370, 12\},\\
\{371, 24\}, \{372, 6\}, \{373, 12\}, \{374, 42\}, \{376, 12\}, \{377, 18\}, \{378, 24\}, \{379, 9\},
\{381, 30\}, \{382, 12\}, \{384, 3\}, \{385, 12\}, \{386, 27\}, \{387, 15\}, \{388, 3\}, \{389, 30\},
\{390, 24\}, \{392, 12\}, \{393, 18\}, \{394, 12\}, \{395, 24\}, \{396, 9\}, \{397, 6\}, \{398, 30\},
\{401, 27\}, \{402, 24\}, \{403, 6\}, \{404, 18\}, \{405, 24\}, \{406, 24\}, \{408, 6\}, \{409, 21\},
\{410, 18\}, \{411, 18\}, \{413, 30\},\\ 
\{414, 30\}, \{416, 6\}, \{417, 18\}, \{418, 12\}, \{419, 27\}, \{420, 12\}, \{421, 12\}, \{422, 15\},\\
\{424, 6\}, \{425, 33\}, \{426, 36\}, \{427, 6\}, \{428, 9\},
\{429, 24\}, \{430, 18\}, \{432, 4\}, \{433, 15\}, \{434, 36\}, \{435, 12\}, \{436, 6\}, \{437, 30\},
\{438, 12\}, \{440, 18\}, \{441, 33\}, \{442, 6\}, \{443, 15\}, \{445, 6\}, \{446, 48\}, \{449, 27\},
\{450, 27\}, \{451, 18\}, \{452, 9\}, \{453, 18\}, \{454, 21\}, \{456, 12\}, \{457, 9\}, \{458, 36\},
\{459, 24\}, \{460, 6\}, \{461, 42\}, \{462, 12\}, \{464, 6\}, \{465, 24\}, \{466, 9\}, \{467, 21\},
\{468, 12\}, \{469, 24\}, \{470, 30\}, \{472, 9\}, \{473, 18\}, \{474, 30\}, \{475, 15\},\\
\{477, 24\}, \{478, 12\}, \{480, 6\}, \{481, 18\}, \{482, 27\}, \{483, 12\}, \{484, 9\}, \{485, 24\}, \{486, 39\},
\{488, 12\}, \{489, 30\}, \{490, 18\}, \{491, 27\}, \{492, 6\}, \{493, 12\}, \{494, 42\}, \{497, 36\},\\
\{498, 12\}, \{499, 9\}, \{500, 12\}.}

\end{document}